# Distillation of photon entanglement using a plasmonic metamaterial

Motoki Asano[1], Muriel Bechu[2,3], Mark Tame[4,5], Şahin Kaya Özdemir [6*], Rikizo Ikuta[1], Durdu Ö. Güney[7], Takashi Yamamoto[1], Lan Yang[6], Martin Wegener[2,3*] and Nobuyuki Imoto[1*]

[1]Department of Material Engineering Science, Graduate School of Engineering Science, Osaka University, Toyonaka, Osaka 560-8531, Japan.

[2]Institute of Applied Physics, Karlsruhe Institute of Technology (KIT), 76128 Karlsruhe, Germany.

[3]Institute of Nanotechnology, Karlsruhe Institute of Technology (KIT), 76128 Karlsruhe, Germany.

[4]School of Chemistry and Physics, University of KwaZulu-Natal, Durban 4001, South Africa.

[5]National Institute for Theoretical Physics, University of KwaZulu-Natal, Durban 4001, South Africa.

[6]Department of Electrical and Systems Engineering, Washington University, St. Louis, MO 63130, USA.

[7]Department of Electrical and Computer Engineering, Michigan Technological University, Houghton, MI 49931, USA.

*Correspondence to: imoto@mp.es.osaka-u.ac.jp, ozdemir@ese.wustl.edu, martin.wegener@kit.edu

**Plasmonics is a rapidly emerging platform for quantum state engineering with the potential for building ultra-compact and hybrid optoelectronic devices. Recent experiments have shown that despite the presence of decoherence and loss, photon statistics and entanglement can be preserved in single plasmonic systems. This preserving ability should carry over to plasmonic metamaterials, whose properties are the result of many individual plasmonic systems acting collectively, and can be used to engineer optical states of light. Here, we report quantum state filtering, also known as 'entanglement distillation' using a metamaterial. We show that the metamaterial can be used to distill highly entangled states from less entangled states. As the metamaterial can be integrated**

**with other optical components this work opens up the intriguing possibility of incorporating plasmonic metamaterials in on-chip quantum state engineering tasks.**

# Introduction

Entanglement plays a key role in a wide variety of quantum information processing tasks[1], enabling quantum communication protocols such as quantum key distribution[2] and quantum computing algorithms providing massive computational speedup compared to conventional computers[3-6]. From a fundamental perspective, entanglement is also at the heart of many foundational quantum phenomena[7]. The task of carrying out filtering operations to improve the amount of entanglement in non-ideal generated states is therefore of great importance in quantum information processing and in studies of fundamental quantum physical effects. Photonic systems in particular represent a flexible test-bed for developing quantum technologies and probing deeper into the foundations of quantum theory[8]. Previous work on photonic entanglement filtering, also called entanglement distillation[9], used standard bulk optical components[10]. Here, we explore the possibility of using metamaterials for this vital task. Metamaterials have recently emerged as highly versatile systems for controlling the behavior of light[11-14]. They are made up of regularly spaced subwavelength components that react collectively to a given optical field in order to elicit a bulk optical response. The use of plasmonic nanostructures for photonic metamaterials is a natural choice due to their electric and magnetic resonances falling within the optical domain[11]. A wide range of applications of plasmonic metamaterials for the optical sciences have been demonstrated so far in the classical regime, including the use of negative refractive index materials[16-19] for superlensing and nano-imaging[20,21], transformation optics[22] and sensing[23]. In the quantum regime, less is known about plasmonic metamaterials[24] and theoretical studies have so far looked at achieving a negative refractive index by manipulating quantum emitters[25], as well as the incorporation of metamaterials with waveguides for reducing the impact of loss in quantum state transfer[26] and entanglement generation[27]. Experimental studies, on the other hand, have focused on basic quantum state transfer effects[28,29], absorption of single photons[30] and quantum interference effects[31]. Most recently the use of 2-dimensional metamaterials, known as metasurfaces[32-36], has gained considerable attention from the metamaterial community due to their ease of fabrication and overall compactness. In this work we explore the use of 2-dimensional plasmonic metamaterials for their potential in quantum state engineering and more specifically the distillation of entanglement. These 2-dimensional metamaterials can be expected to be more readily accessible than their 3-dimensional counterparts for realizing advanced quantum applications in the near future. Our study builds upon previous work on the classical characterization of the collective response of nanostructured arrays[37,38], and in the quantum regime on the assisted-transmission of entanglement in periodic plasmonic nano-hole arrays[39], and the remote control of transmission of single photons[40]. However, different to these works, here we go beyond a simple transmission scenario in the quantum regime and show that plasmonic nanostructured arrays can be used not only for basic transfer of quantum information, but also for the manipulation of quantum information in the form of quantum state engineering. Furthermore, we have fully characterized the metamaterial nanostructured arrays using the rigorous technique of quantum process tomography, showing how to characterize the optical response of metamaterials in the quantum regime.

# Results

The task of entanglement distillation refers to the process of extracting a smaller number of highly entangled states from an ensemble of less-entangled states[9]. Entanglement shared between two parties (bi-partite entanglement) is the simplest form of entanglement. A two-qubit state encoded in the polarization degrees of freedom of two photons (each in a spatially separate path) of the form

$$|\Phi_\varepsilon\rangle = \tfrac{1}{\sqrt{1+\varepsilon^2}}(\varepsilon|H\rangle|H\rangle + |V\rangle|V\rangle) , \qquad (1)$$

where $|H\rangle$ and $|V\rangle$ represent the horizontal and vertical polarization state of a photon, is a non-maximally entangled pure state for $\varepsilon \neq 1$. It can be transformed into a maximally entangled state (a Bell state) of the form $|\Phi^+\rangle = (|H\rangle|H\rangle + |V\rangle|V\rangle)/\sqrt{2}$ by using a local operator, acting on only one of the photons, that induces a polarization dependent modification of the amplitudes. In order to realize this operation we utilize polarization dependent extinction introduced by the collective action of many plasmonic resonators in a metamaterial.

The metamaterial used in our experiments consists of an assembly of gold nanoantennas grown on an ITO-coated suprasil substrate, as described in the Supplementary Material. The final structure represents an array of straight nanoantennas occupying a footprint of up to $10^{-4}$ cm$^2$, as shown in Fig. 1a. The dimensions of the rod-like nanoantennas are 95-110 nm in length, 39 nm in width and 30 nm in height, with a spacing of 200 nm center-to-center between them, thus achieving a nanorod density of ~$10^9$ cm$^{-2}$. The dimensions and the spacing of the antennas are much smaller than the wavelength of the photons used in our experiments (790 nm), so only average values of nanorod assembly parameters are important, and individual nanorod size deviations have no influence on the optical properties that are well described by an effective medium model[32-36]. When V-polarized light impinges onto vertical metallic nanoantennas of a certain length a plasmonic resonance is excited in the form of light coupled to a collective oscillation of free electrons in the conduction band – a localized surface plasmon (LSP). The generation of the LSP leads to a dip in the transmission spectra of the light at the resonant frequency. This dip reflects the fact that some of the light is reflected back into the far field and some is absorbed by the LSP. Due to the Ohmic resistance faced by the oscillating electrons, the energy used to excite them is partly dissipated, the amount by which depends on the dimensions of the nanoantenna. The dimensions of the nanoantenna also determine the resonant frequency of the LSP and therefore the position of the dip in the overall transmission spectrum. On the other hand, light polarized perpendicular (H-polarized) to the antennas does not excite the plasmonic resonance and passes the sample almost unchanged. Fig. 1b depicts the transmission spectra of two typical nanoantenna array metamaterials used in our experiments. A clear polarization dependence of the transmission spectra is seen.

In our setup (Fig. 1c), we prepared polarization entangled photon pairs at a wavelength of $\lambda$=790 nm via spontaneous parametric down conversion (SPDC) in Type-I phase matched nonlinear crystals (β-barium borate, BBO) stacked together such that their optical axes are orthogonal to each other[41]. The SPDC pump laser with a wavelength of λ=395 nm is obtained by frequency doubling the light from a mode locked Ti: Sapphire laser at λ=790 nm. We arbitrarily set ε of the entangled state by varying the polarization of the pump laser[41], *i.e.* when the polarization of the pump was set to diagonal polarization the prepared photon pair was maximally entangled (ε = 1), whereas when it was set to horizontal polarization the prepared photons were in the product state |VV⟩ (ε = 0). The difference in the group velocity of photons with different polarization was compensated by birefringent crystals (BC) and the phase between horizontal and vertical polarization was adjusted by a set of quartz crystals represented as PS.

We performed a series of experiments by inserting different metamaterial samples (with different lithography parameters – hence different nanoantenna resonance positions) into the optical path of one of the photons of the entangled photon pair. One photon was transmitted through the metamaterial after which it and the other photon of the pair were sent to independent single-mode-fiber-coupled silicon avalanche photo diodes (APDs). Before being coupled into fibers the photons passed through interference filters of bandwidth 2.7 nm, and a series of a half-wave plate (HWP), a quarter-wave plate (QWP) and a polarizing beamsplitter (PBS) placed on their respective paths. The interference filter and single mode fiber performed the selection of the spectral and spatial mode of the photons respectively. The HWP, QWP and PBS were used to choose the measurement basis states |H⟩, |V⟩, |D⟩ = (|H⟩ + |V⟩)/$\sqrt{2}$ and |R⟩ = (|H⟩ + i|V⟩)/$\sqrt{2}$ required for the characterization of the final states using quantum state tomography[42] (QST). The spot size of the beam on the nanoantenna array of the metamaterial was adjusted to be ~90 μm in diameter to ensure the collective electromagnetic response of the nanoantennas (~$10^6$ nanoantennas in the beam path). We positioned the different metamaterial samples such that the vertical polarization of the photons was parallel to the long-axis of the nanoantennas.

In the first set of experiments, we performed quantum process tomography[43,44] (QPT) to characterize the nanoantenna arrays used in our experiments. QPT allows us to reconstruct the action of the metamaterial on the polarization state of a single photon as an effective quantum channel. To reconstruct the channel we probe the metamaterial with different photonic probe states. For this purpose, we set the pump laser to H polarization so that two photons with V-polarization are prepared by SPDC in one of the BBO crystals. We then insert a HWP and a QWP in front of the metamaterial sample to prepare the first photon in one of the four probe states $|H\rangle$, $|V\rangle$, $|D\rangle$ and $|R\rangle$ required for QPT. The HWP and QWP on the path of the second photon were set such that V polarized photons are always detected by the APD. The detection of a photon in the second path heralds the presence of a single probe photon in the first path. The photons in the probe states in the first path were sent to the metamaterial and QST was performed on the ones that were transmitted through the metamaterial by recording the coincidence events, *i.e.* when APDs in the first and second paths detect a photon at the same time. From the collected experimental data, we reconstructed the single-photon process matrices, known as $\chi$ matrices, for seven different metamaterial nanoantenna arrays. The $\chi$ matrices obtained for two of the nanoantenna arrays are shown in Fig. 2 (see the Supplementary Material for all $\chi$ matrices). We found that the $\chi$ matrices of the nanoantenna arrays are well described by the $\chi$ matrix of a partial polarizer represented by a single Kraus operator $K_0 = |H\rangle\langle H| + \sqrt{T_V}|V\rangle\langle V|$ corresponding to a non-trace preserving channel[45], *i.e.* $\rho \to K_0 \rho K_0^\dagger$, where $\rho$ is the input single-photon state in the polarization basis. This photonic channel is equivalent to the general form $\rho \to \sum_{ij} \chi_{ij} E_i \rho E_j^\dagger$, where the single-qubit Pauli operators, $E_i = I$, X, Y and Z, provide a complete basis for the Hilbert space and the elements of the $\chi$ matrix are chosen to match the action of $K_0$ (see Supplementary Material). In order to quantify how close the metamaterial samples are to an ideal partial polarizer model we calculated the process fidelity $F_P(T_V) = \text{Tr}\left(\sqrt{\sqrt{\chi}\chi_{id}\sqrt{\chi}}\right)^2 / \text{Tr}(\chi)\text{Tr}(\chi_{id})$ of the two $\chi$ matrices shown in Fig. 2 to an ideal partial polarizer $\chi_{id}$. In general, the fidelity ranges from 0 to 1, with 1 corresponding to a complete match for the channels. We find process fidelities of $0.93 \pm 0.01$ and $0.90 \pm 0.01$ by maximization over $T_V$, which yielded $T_V = 0.11 \pm 0.01$ and $T_V = 0.41 \pm 0.01$, respectively. The $T_V$ values obtained from QPT agree well with the measured $T_V$ values using classical FTIR (see Fig. 1b). These results confirm that the plasmonic metamaterial fabricated with different nanoantenna array parameters has a polarization dependent transmission in the low-light intensity quantum regime and can therefore be used to induce a collective polarization dependent loss at the single-photon level.

Next, we performed experiments to demonstrate that our plasmonic metamaterial can be used to distill highly entangled pure states from an ensemble of less-entangled pure states. First, we generated the initial less-entangled pure state given in Eq. (1) by varying the polarization of the pump in order to set the value of $\varepsilon$, and checked the entanglement distillation performance of each of the nanoantenna arrays. As a control experiment, we sent one of the photons of the prepared entangled state to a portion of the metamaterial sample where there were no nanoantennas, *i.e.* the photon passes through the glass substrate only, and performed QST of the two photons arriving at the APDs. The reconstructed density matrix of this initial state is given in Fig. 3a. We estimate the purity of this state as 0.97±0.01 using $\text{Tr}(\rho^2)$, with a value of 1 corresponding to a completely pure state[44], and subsequently calculate the value of $\varepsilon$ as $\varepsilon_{\text{exp}} = 0.49 \pm 0.02$ using $\varepsilon_{\text{exp}} \equiv \text{Tr}[\rho|HH\rangle\langle HH|]/\text{Tr}[\rho|VV\rangle\langle VV|]$, where $\rho$ is the density operator of the state obtained from QST. The fidelity of this initial state with respect to the non-maximally entangled state with $\varepsilon = 0.49$ is $0.96 \pm 0.01$ using $F = \langle \Phi_\varepsilon|\rho|\Phi_\varepsilon \rangle$ and the fidelity with respect to the maximally entangled state $(|HH\rangle + |VV\rangle)/\sqrt{2}$ (with $\varepsilon = 1$) is $0.85 \pm 0.01$. We also calculated the entanglement of formation[1] (EOF) that quantifies the amount of entanglement in the generated bipartite state as 0.66±0.01, verifying its non-maximal value of entanglement.

After confirming the purity and the amount of entanglement of this initial non-maximally entangled state, we performed experiments with the state using the seven different metamaterial nanoantenna arrays. Figure 3b presents the reconstructed density matrix of the distilled state that had the highest EOF observed in our experiments. This distilled state has a fidelity of 0.95±0.01 with respect to a maximally entangled state and an EOF of 0.93±0.02. The density matrices of the distilled states obtained with the seven different nanoantenna arrays are given in the Supplementary Material. In Fig. 3c, we show the EOF, fidelity and purity of the distilled states for the seven nanoantenna arrays used in the experiments, confirming the applicability of these metamaterial arrays for distilling highly entangled states from less-entangled starting states. The purity of the output states keeps a constant high value (close to 0.95), which reflects the preservation of the coherence of the photons during the filtering process.

We also tested the performance of a fixed metamaterial nanoantenna array for entanglement distillation of different initial states of the form $|\Phi_\varepsilon\rangle$ and $|\Psi_\varepsilon\rangle = (\varepsilon|H\rangle|V\rangle + |V\rangle|H\rangle)/\sqrt{1+\varepsilon^2}$. The results are shown in Fig. 3d which shows that when a fixed nanoantenna array is used, the fidelity and the EOF of the distilled state depend on the value of $\varepsilon$ for the initial state, and that there is an $\varepsilon$ value for which the specific array is optimal for entanglement distillation.

Next, we tested the ability of the local filtering process of the metamaterial nanoantenna arrays to distill entangled states with a higher amount of entanglement from partially mixed states containing lower amounts of entanglement. In order to prepare an entangled state of a partially mixed state, we placed a quartz crystal (12.8 mm thick) inserted between two HWPs in front of the metamaterial sample, as shown in Fig. 1c. Due to the group velocity difference between H and V polarizations, the quartz crystal partially destroys the coherence, resulting in the partially mixed state. We control the degree of decoherence by rotating the first HWP to prepare arbitrary superposition of H- and V-polarizations. The HWP after the quartz crystal is used to rotate the polarization back to the initial polarization basis. By using this technique, we prepared three different non-maximally entangled partially mixed states of the form $\rho_{\varepsilon,\lambda} = \frac{1}{1+\varepsilon^2}\left[\varepsilon^2|HH\rangle\langle HH| + |VV\rangle\langle VV| + \varepsilon\left(1-\frac{\lambda}{2}\right)(|HH\rangle\langle VV| + |VV\rangle\langle HH|)\right]$ and three of the form $\sigma_{\varepsilon,\lambda} = \frac{1}{1+\varepsilon^2}\left[\varepsilon^2|HV\rangle\langle HV| + |VH\rangle\langle VH| + \varepsilon\left(1-\frac{\lambda}{2}\right)(|HV\rangle\langle VH| + |VH\rangle\langle HV|)\right]$ as starting states (see Supplementary Material) and performed the distillation process using a fixed metamaterial nanoantenna array. In Fig. 4, we present the density matrices of two of the initial mixed states and the final distilled states obtained from the metamaterial (see Supplementary Material for density matrices of the other four mixed states). From the tomographically reconstructed density matrix of each of the initial and distilled states, we estimated the fidelity and EOF (see Table 1). These values clearly show that the distilled states have a higher entanglement and a higher fidelity than the starting states. Table 1 also includes the estimated values of $\varepsilon$ and $\lambda$ before and after the distillation.

We should emphasize here that the filtering process and coincidence detection select a particular subensemble from the ensemble of the starting initial states, with coincidence detection rates before and after filtering corresponding to 4490 and 1823 counts per second, respectively. The amount of entanglement in the states in the selected subensemble is higher than the amount of entanglement of the larger ensemble containing the initial states. The unselected states have much lower entanglement. This does not contradict with the fact that entanglement of an ensemble of states cannot be increased by LOCC. That is, if we consider all the selected and unselected states the average entanglement does not increase. The metamaterial thus enables a quantum selection process to take place so that all of the partially entangled states can be distilled into a smaller number of higher entangled states that may then be used for further quantum information processing tasks.

## Discussion

Our experiment demonstrates that plasmonic metamaterials can be used for a quantum information processing task in the form of the distillation of quantum entanglement. This clearly shows that an array of nanostructures in a metamaterial can be used to perform quantum state engineering. Our work goes beyond previous works in plasmonics and metamaterials where the initial interest was to show that quantum features of plasmons are similar to those of photons and that they are preserved during the photon-plasmon-photon interconversion process[24]. Another key difference of our work is that it relies on the collective response of many subwavelength plasmonic structures within the plasmonic metamaterial, which is in stark contrast to most other studies where the quantum response of only single plasmonic structures has been studied. Due to the 2-dimensional nature of the metamaterial investigated, the nanoantenna structures can be fabricated with well-controlled dimensions, providing a high quality design with a small-lateral footprint. This makes it ideal for integration with wavelength-scale plasmonic[46] and dielectric[47] components, such as on-chip optical waveguides, where it could be used for entanglement distillation and other quantum information processing tasks. Future work on developing tunable nanoantenna structures could lead to 2-dimensional metamaterials that provide enhanced functionality for entanglement distillation and other quantum state engineering tasks by enabling one to tune the metamaterial response for optimum performance.

**Acknowledgements** This work was supported by JSPS Grant-in-Aid for Scientific Research (A) 25247068, (B) 15H03704 and (B) 26286068; The Karlsruhe School of Optics & Photonics (KSOP); The South African National Research Foundation; The South African National Institute for Theoretical Physics.

**Figure 1. The plasmonic metamaterial and experimental setup used for entanglement distillation.** (a) An illustration of the metamaterial illuminated by a laser beam together with the SEM image. The metamaterial was fabricated on an ITO-coated suprasil substrate by exposing a positive tone photoresist by electron-beam, which was then developed, leaving a mask. Subsequent gold evaporation and lift-off yielded the gold nanoantennas with typical dimensions of 112 nm × 39 nm × 30 nm. (b) Transmission spectra obtained for two different gold nanoantenna arrays. Solid and filled points belong to the different nanoantenna arrays. Boxes and circles correspond to horizontally (H-) and vertically (V-) polarized coherent light, respectively. The antennas have close-to-unity transmission for H-polarized light at around ~ 790 nm (dashed line) where the V-polarized light has low transmission on resonance. (c) An illustration of the experimental setup. See main text for details. HWP: Half-wave plate, QWP: Quarter-wave plate, BBO: β-barium borate crystal, IF: Interference filter, PBS: Polarizing beamsplitter, APD: Avalanche photodiode. The optical components in the 'decoherence' box are used to prepare non-maximally entangled mixed states.

**Figure 2. Characterization of the metamaterial by quantum process tomography.** Experimentally obtained process matrices (χ matrices) for two different metamaterials used in the experiments for entanglement distillation. The process matrices are given in the basis defined by the single-qubit Pauli operators, $E_i =$ I, X, Y and Z, where a single qubit is modified as $\rho \to \sum_{ij} \chi_{ij} E_i \rho E_j^\dagger$. (a) Real part of the process matrix for metamaterial sample 1 (left) and an ideal partial polarizer with $T_V = 0.11 \pm 0.01$ (right). (b) Imaginary part of the process matrices for the cases considered in panel a. (c) Real part of the process matrix for metamaterial sample 2 (left) and an ideal partial polarizer with $T_V = 0.41 \pm 0.01$ (right). (d) Imaginary part of the process matrices for the cases considered in panel c. The process fidelities of the metamaterial samples to the ideal partial polarizer cases given are 0.93±0.01 ( Tr(χ)=0.53±0.01 ) and 0.90±0.01 ( Tr(χ)=0.69±0.01 ). See Supplementary Material for χ matrices of the other five nanoantenna arrays used in the experiments.

**Figure 3. Distillation of highly entangled states from non-maximally entangled pure states using metamaterial nanoantenna arrays.** (a) Density matrix of the initial state of the form $|\Phi_\varepsilon\rangle = (\varepsilon|H\rangle|H\rangle + |V\rangle|V\rangle)/\sqrt{1+\varepsilon^2}$. (b) Density matrix of the metamaterial distilled state. Note that the weights of the components in the distilled state are more balanced than the starting state. (c) Entanglement distillation performance of different metamaterials for a fixed non-maximally entangled pure state. Entanglement of formation (EOF) (red), fidelity (green) and purity (blue). The EOF and the fidelity of the distilled states with respect to the maximally entangled state are higher than the initial state (no array case) for all tested metamaterial nanoantenna arrays. The antenna arrays do not affect the purity of the state. (d) Entanglement distillation performance of a fixed metamaterial nanoantenna array for various non-maximally entangled pure states of the form $|\Phi_\varepsilon\rangle = (\varepsilon|H\rangle|H\rangle + |V\rangle|V\rangle)/\sqrt{1+\varepsilon^2}$ (blue) and $|\Psi_\varepsilon\rangle = (\varepsilon|H\rangle|V\rangle + |V\rangle|H\rangle)/\sqrt{1+\varepsilon^2}$ (red). The inset shows the fidelity of the distilled state to the maximally entangled state $|\Phi_{\varepsilon=1}\rangle$ and $|\Psi_{\varepsilon=1}\rangle$ respectively.

**Figure 4. Distillation of highly entangled states from non-maximally entangled partially mixed states using metamaterial nanoantenna arrays.** (a) Density matrix of the starting mixed state of the form $\rho_{\varepsilon,\lambda} = \frac{1}{1+\varepsilon^2}\left[\varepsilon^2|HH\rangle\langle HH| + |VV\rangle\langle VV| + \varepsilon\left(1-\frac{\lambda}{2}\right)(|HH\rangle\langle VV| + |VV\rangle\langle HH|)\right]$. (b) Density matrix of the distilled state for the starting mixed state of a . (c) Density matrix of the starting mixed state of the form $\sigma_{\varepsilon,\lambda} = \frac{1}{1+\varepsilon^2}\left[\varepsilon^2|HV\rangle\langle HV| + |VH\rangle\langle VH| + \varepsilon\left(1-\frac{\lambda}{2}\right)(|HV\rangle\langle VH| + |VH\rangle\langle HV|)\right]$. (d) Density matrix of the distilled state for the starting mixed state of c. See Table 1 for the estimated EOF, fidelity and purity of the starting states and distilled states. See Methods for the density matrices of all tested mixed states.

**Table 1. Summary of the distillation data for non-maximally entangled partially mixed states.** The table shows the measured fidelity, EOF and the estimated values of ε and λ parameters of the initial and the distilled states. The errors are calculated from a Monte Carlo simulation assuming Poisson statistics. The starting states labeled from 1 to 3 are of the form $\rho_{\varepsilon,\lambda} = \frac{1}{1+\varepsilon^2}\left[\varepsilon^2|HH\rangle\langle HH| + |VV\rangle\langle VV| + \varepsilon\left(1-\frac{\lambda}{2}\right)(|HH\rangle\langle VV| + \right.$

$|VV\rangle\langle HH|)\Big]$, and those from 3 to 6 are of the form $\sigma_{\varepsilon,\lambda} = \frac{1}{1+\varepsilon^2}\Big[\varepsilon^2|HV\rangle\langle HV| + |VH\rangle\langle VH| + \varepsilon\left(1 - \frac{\lambda}{2}\right)(|HV\rangle\langle VH| + |VH\rangle\langle HV|)\Big]$.

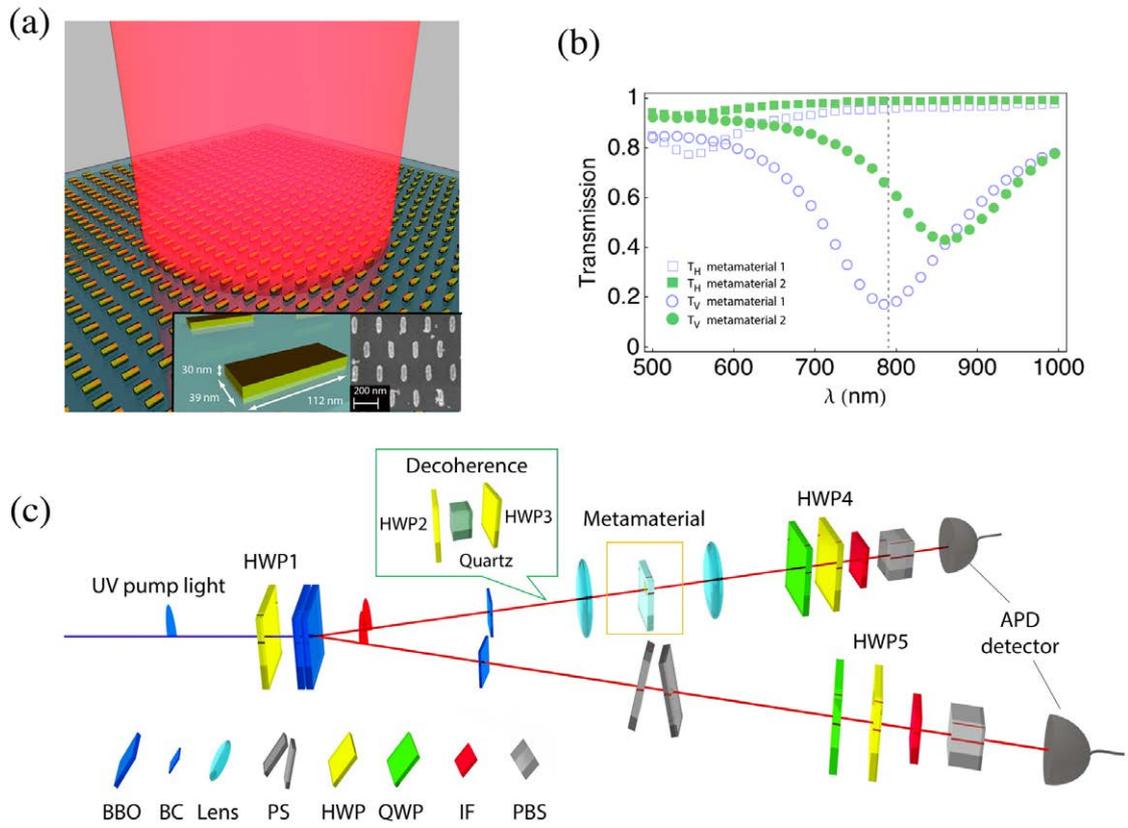

**Figure 1**

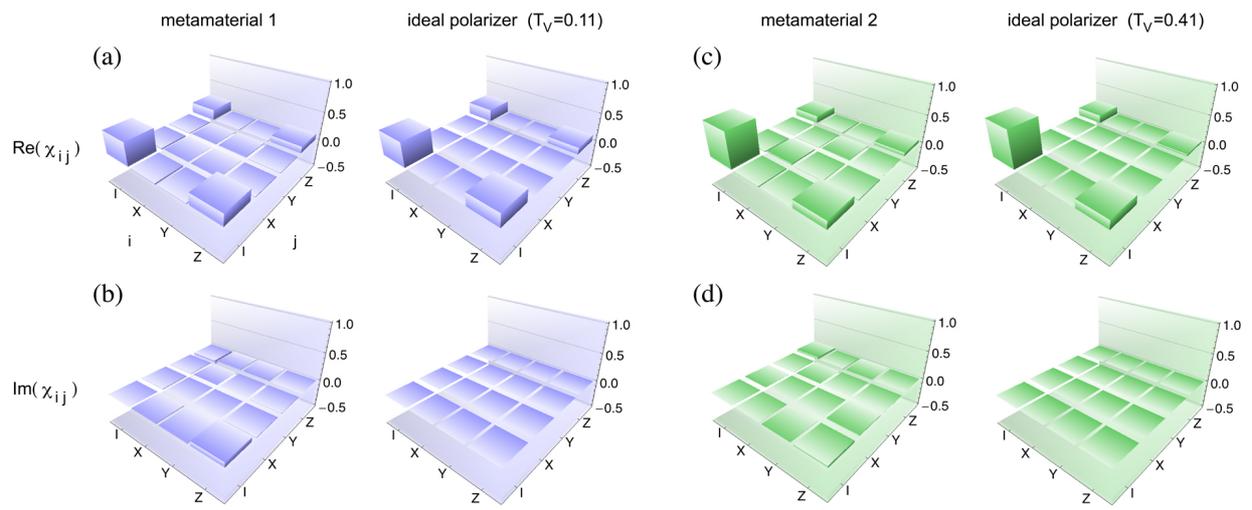

**Figure 2**

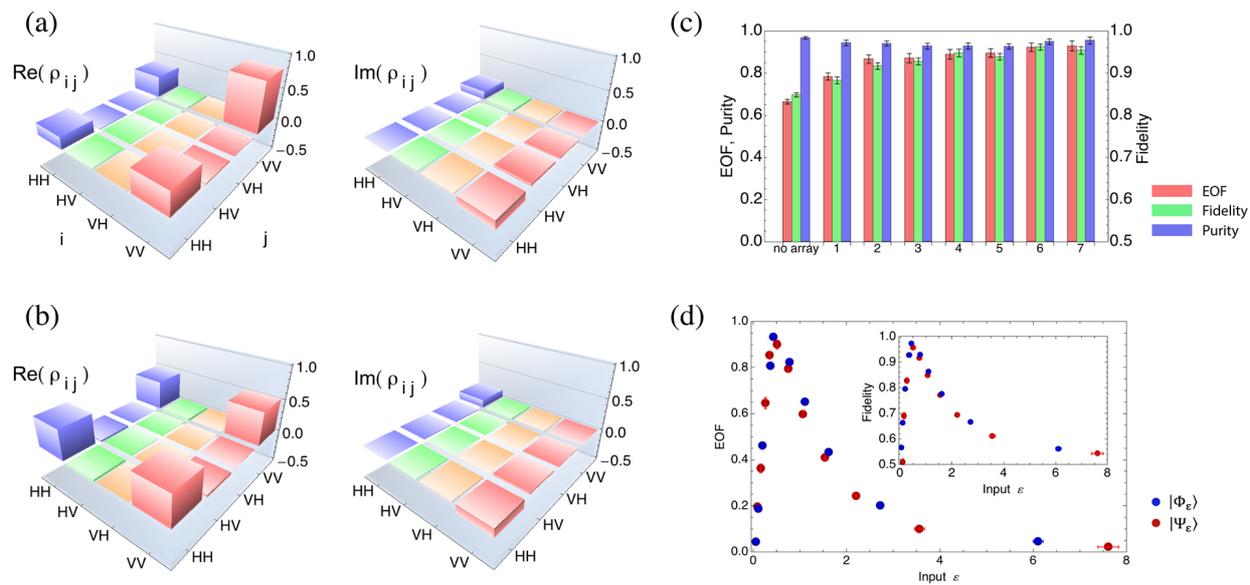

**Figure 3**

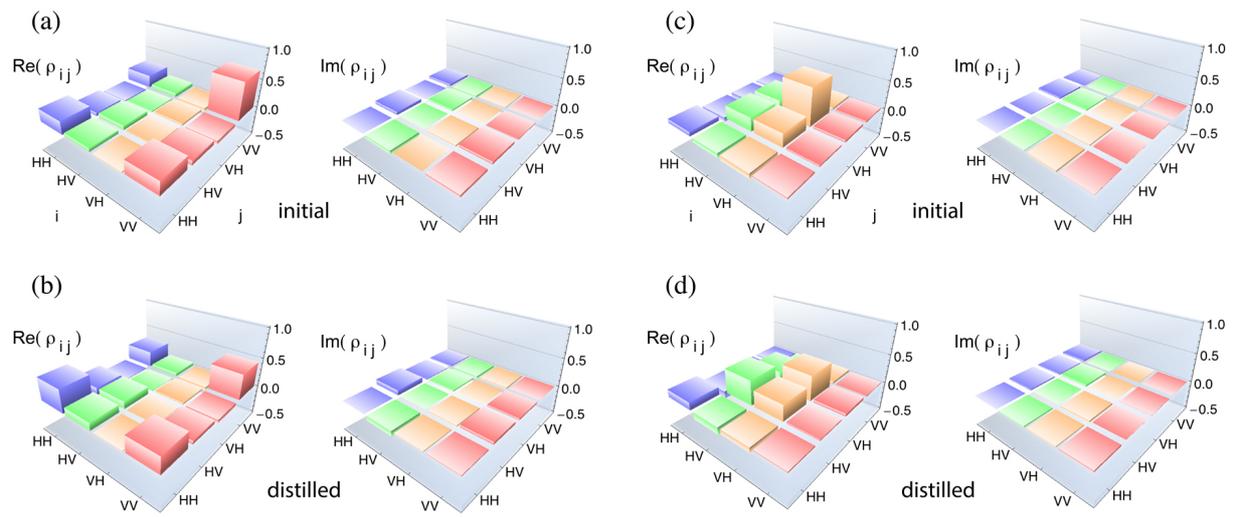

Figure 4

| State | Initial ε | Initial λ | Initial Fidelity | Initial EOF | Distilled ε | Distilled λ | Distilled Fidelity | Distilled EOF |
|---|---|---|---|---|---|---|---|---|
| 1 | 0.59±0.01 | 0.54 ±0.05 | 0.80±0.02 | 0.50±0.03 | 0.96±0.01 | 0.51±0.04 | 0.84±0.01 | 0.60± 0.03 |
| 2 | 0.58±0.01 | 0.65 ±0.05 | 0.74±0.02 | 0.38±0.04 | 0.94±0.01 | 0.52±0.04 | 0.78±0.01 | 0.50±0.03 |
| 3 | 0.59±0.01 | 0.83±0.05 | 0.67±0.02 | 0.25±0.03 | 0.96±0.01 | 0.69±0.04 | 0.70±0.01 | 0.35±0.03 |
| 4 | 0.61± 0.01 | 0.49 ±0.04 | 0.82±0.01 | 0.54±0.04 | 0.98±0.01 | 0.43±0.03 | 0.87±0.01 | 0.66± 0.03 |
| 5 | 0.60±0.01 | 0.61±0.04 | 0.76±0.01 | 0.43±0.03 | 0.98±0.01 | 0.58±0.03 | 0.80±0.01 | 0.54±0.03 |
| 6 | 0.60 ±0.00 | 0.81±0.04 | 0.68±0.01 | 0.28±0.03 | 1.00±0.01 | 0.69±0.03 | 0.70±0.01 | 0.38±0.02 |

**Table 1**

# Supplementary Information: Distillation of photon entanglement using a plasmonic metamaterial

Motoki Asano[1], Muriel Bechu[2,3], Mark Tame[4,5], Şahin Kaya Özdemir[6*], Rikizo Ikuta[1], Durdu Ö. Güney[7], Takashi Yamamoto[1], Lan Yang[6], Martin Wegener[2,3*] and Nobuyuki Imoto[1*]

[1]*Department of Material Engineering Science, Graduate School of Engineering Science, Osaka University, Toyonaka, Osaka 560-8531, Japan.*

[2]*Institute of Applied Physics, Karlsruhe Institute of Technology (KIT), 76128 Karlsruhe, Germany.*

[3]*Institute of Nanotechnology, Karlsruhe Institute of Technology (KIT), 76128 Karlsruhe, Germany.*

[4]*School of Chemistry and Physics, University of KwaZulu-Natal, Durban 4001, South Africa.*

[5]*National Institute for Theoretical Physics, University of KwaZulu-Natal, Durban 4001, South Africa.*

[6]*Department of Electrical and Systems Engineering, Washington University, St. Louis, MO 63130, USA.*

[7]*Department of Electrical and Computer Engineering, Michigan Technological University, Houghton, MI 49931, USA.*

*Correspondence to: imoto@mp.es.osaka-u.ac.jp, ozdemir@ese.wustl.edu, martin.wegener@kit.edu*

## 1. Metamaterial fabrication

The metamaterials were fabricated by electron-beam lithography followed by a lift-off procedure. Precisely, 5 mm × 5 mm suprasil substrates were prepared for electron-beam lithography by depositing a 5 nm thin layer of indium tin oxide (ITO) by electron-beam evaporation. Next, an approximately 200 nm thick film of polymethylmetacrylate photoresist (MicroChem) was spin-coated on top of the ITO. This positive tone photoresist was then patterned by electron-beam writing (Raith e-line). We wrote 160 different arrays of straight nano-antennas. The nano-antenna length and thickness was varied between the arrays to shift the spectral position of the resonance. Each array has a footprint of 100 μm × 100 μm and contains antennas with a horizontal spacing of 200 nm and nominal lengths between 95 nm and 110 nm. After exposure, the samples were developed in a 1:3 solution of methyl isobuthyl ketone (MIBK) and isopropanol. This process dissolves the long-chained molecules of the photoresist that have been broken up during the exposure process and thus creates a mask. Onto this mask, 30 nm of gold were deposited by high-vacuum electron-beam evaporation. To lift off the PMMA mask and the excess gold, the samples were exposed to a bath of Allresist remover AR 300-70 at 50° C until the lift-off was completed.

**2. Quantum process tomography**

The general form of a quantum channel corresponding to a completely positive map on the state ρ is given by $\rho \to \sum_{ij} \chi_{ij} E_i \rho E_j^\dagger$, where $\sum_i E_i^\dagger E_i \leq I$ (with equality for a trace-preserving map). For a single qubit, the Pauli operators, $E_i = I, X, Y$ and $Z$, provide a complete basis for the Hilbert space. For the ideal model of a partial polarizer given in the main text we have the channel $\rho \to K_0 \rho K_0^\dagger$, where the Kraus operator $K_0 = |H\rangle\langle H| + \sqrt{T_V}|V\rangle\langle V|$ corresponds to a non-trace preserving channel. The equivalent form of this channel in the Pauli basis is given by the general map $\varepsilon: \rho \to \sum_{ij} \chi_{ij} E_i \rho E_j^\dagger$, where the elements of the χ matrix are

$$\chi_{id} = \begin{pmatrix} (1 + 2\sqrt{T_V} + T_V)/4 & 0 & 0 & (1 - T_V)/4 \\ 0 & 0 & 0 & 0 \\ 0 & 0 & 0 & 0 \\ (1 - T_V)/4 & 0 & 0 & (1 - 2\sqrt{T_V} + T_V)/4 \end{pmatrix} \quad (S1)$$

which gives $\text{Tr}(\chi) = (1 + T_V)/2$. This channel is trace preserving (and unitary) only for $T_V = 1$. In order to obtain the elements of an experimental χ matrix for a given single-qubit channel ε, one can probe it with the four states $|H\rangle, |V\rangle, |D\rangle$ and $|R\rangle$, which allow the reconstruction of the action of ε on the different elements of an arbitrary input state: $\varepsilon(|H\rangle\langle H|)$, $\varepsilon(|V\rangle\langle V|)$, $\varepsilon(|H\rangle\langle V|) = \varepsilon(|D\rangle\langle D|) + i\varepsilon(|R\rangle\langle R|) - \frac{1}{2}(1 + i)[\varepsilon(|H\rangle\langle H|) + \varepsilon(|V\rangle\langle V|)]$ and $\varepsilon(|V\rangle\langle H|) = \varepsilon(|D\rangle\langle D|) - i\varepsilon(|R\rangle\langle R|) - \frac{1}{2}(1 - i)[\varepsilon(|H\rangle\langle H|) + \varepsilon(|V\rangle\langle V|)]$. From this it is straightforward to extract out the χ matrix elements [1]. To obtain the different probe state outputs $\varepsilon(|i\rangle\langle i|)$, we prepare each of the probe states and send them into the metamaterial. The output states are then obtained from quantum state tomography. Note, however, that the channel is expected to be non-trace preserving. Thus we must weight the different output states by their relative probability of being produced, given a probe state was input to the channel. For instance, the probe state $|V\rangle$ is only expected to be transmitted through the metamaterial with probability $T_V$ in the ideal case, thus the output state $\varepsilon(|V\rangle\langle V|) = |V\rangle\langle V|$ would be produced with probability $T_V$ and any channel reconstruction would need to weight $\varepsilon(|V\rangle\langle V|)$ by the factor $T_V$. More generally, for a fixed time period we count the number of output states transmitted by a given input probe state when there is no metamaterial present (glass substrate only). This is obtained by measuring the total number of counts for measurements in the $|H\rangle/|V\rangle$ basis, providing a reference value, $N_i^r$, for each probe state i when there is no metamaterial (corresponding to the identity operation). In the presence of the metamaterial we again count the number of output states transmitted by a given probe state using the $|H\rangle/|V\rangle$ basis, which provides the value $N_i$. The relative probability of an output state being produced by the metamaterial given a probe state was input is then given by $p_i = N_i/N_i^r$. This weighting of the probe state outputs $\varepsilon(|i\rangle\langle i|)$ leads to a non-trace preserving $\chi$ matrix. In Fig. S1, we show the reconstructed $\chi$ matrices for the seven different metamaterials studied in our experiment. The process fidelity $F_P = \text{Tr}\left(\sqrt{\sqrt{\chi}\chi_{id}\sqrt{\chi}}\right)^2 / \text{Tr}(\chi)\text{Tr}(\chi_{id})$ of each $\chi$ matrix with respect to the ideal partial polarizer $\chi_{id}$ is maximized over the variable $T_V$, leading to the ideal $\chi_{id}$ matrices shown to the right of the corresponding experimentally reconstructed ones. The process fidelities are given in the caption along with the maximized $T_V$ values.

## 3. Density matrices of distilled states for pure input states

In Figure S2 we show the density matrices of the distilled states from each of the seven metamaterials.

## 4. Density matrices of distilled states for non-maximally entangled partially mixed input states

To prepare non-maximally entangled partially mixed states in our experiment, we implement a phase damping channel by using a quartz plate sandwiched between two HWPs inserted into the path of one of the photons. The quartz plate induces phase damping in the polarization basis by introducing a delay between photons with horizontal polarization and those with vertical polarization. This delay is comparable to the coherence time of the two terms in the non-maximally entangled input state but much shorter than the coincidence window and therefore produces an effective phase damping effect. The HWPs enable the amount of phase damping to be controlled by rotating the polarization basis in which the phase damping occurs. The Kraus operators corresponding to this optical configuration are given by

$$E_1(\theta, \lambda) = U_{HWP}^{-1}(\theta) \begin{pmatrix} 1 & 0 \\ 0 & \sqrt{1-\lambda} \end{pmatrix} U_{HWP}(\theta) \otimes I, \tag{S2}$$

$$E_2(\theta, \lambda) = U_{HWP}^{-1}(\theta) \begin{pmatrix} 0 & 0 \\ 0 & \sqrt{\lambda} \end{pmatrix} U_{HWP}(\theta) \otimes I, \tag{S3}$$

where $U_{HWP}(\theta) = \begin{pmatrix} \cos\theta & \sin\theta \\ \sin\theta & -\cos\theta \end{pmatrix}$ and I is the identity operation. Here, each matrix is written in the $|H\rangle/|V\rangle$ basis, $\lambda$ represents the degree of phase damping corresponding to the difference of group velocity between each polarization (due to the quartz plate). Using the Kraus representation, the partially mixed state produced by acting on the non-maximally entangled state $|\Phi_\varepsilon\rangle$ is expressed by [2]

$$\rho_{in}(\theta, \lambda, \varepsilon) = \frac{E_1|\Phi_\varepsilon\rangle\langle\Phi_\varepsilon|E_1^\dagger + E_2|\Phi_\varepsilon\rangle\langle\Phi_\varepsilon|E_2^\dagger}{Tr[E_1|\Phi_\varepsilon\rangle\langle\Phi_\varepsilon|E_1^\dagger + E_2|\Phi_\varepsilon\rangle\langle\Phi_\varepsilon|E_2^\dagger]} = \frac{1}{1+\varepsilon^2} \begin{pmatrix} a_1 & a_2 & a_4 & a_7 \\ a_2 & a_3 & a_5 & a_8 \\ a_4 & a_5 & a_6 & a_9 \\ a_7 & a_8 & a_9 & a_{10} \end{pmatrix} \tag{S4}$$

where $|\Phi_\varepsilon\rangle$ corresponds to the pure state defined in Eq. (1), and each element is exactly calculated as the following:

$$a_1 = \varepsilon^2 \left(1 - 2\cos^2\theta \sin^2\theta \left(1 - \sqrt{1-\lambda}\right)\right) \tag{S5}$$

$$a_2 = \frac{\varepsilon}{4} \sin 4\theta \left(1 - \sqrt{1-\lambda}\right)$$

$$a_3 = 2\cos^2\theta \sin^2\theta \left(1 - \sqrt{1-\lambda}\right)$$

$$a_4 = \frac{\varepsilon^2}{4} \sin 4\theta \left(1 - \sqrt{1-\lambda}\right)$$

$$a_5 = 2\varepsilon \cos^2\theta \sin^2\theta \left(1 - \sqrt{1-\lambda}\right)$$

$$a_6 = \frac{\varepsilon^2}{2}\sin^2 2\theta \left(1 - \sqrt{1-\lambda}\right)$$

$$a_7 = \frac{\varepsilon}{4}\left(1 + 3\sqrt{1-\lambda} - \left(1 - \sqrt{1-\lambda}\right)\cos 4\theta\right)$$

$$a_8 = -\frac{1}{4}\sin 4\theta \left(1 - \sqrt{1-\lambda}\right)$$

$$a_9 = -\frac{\varepsilon}{4}\sin 4\theta \left(1 - \sqrt{1-\lambda}\right)$$

$$a_{10} = 1 - 2\cos^2\theta \sin^2\theta \left(1 - \sqrt{1-\lambda}\right).$$

We can obtain a simple approximate form of the density matrix for the non-maximally entangled partially mixed state by omitting terms that are higher than the second order of $\lambda$ and $\theta$, and cross terms to first order,

$$\rho_{in}(\lambda, \varepsilon) \approx \frac{1}{1+\varepsilon^2}\left[\begin{array}{c}\varepsilon^2|HH\rangle\langle HH| + |VV\rangle\langle VV| \\ +\varepsilon\left(1 - \frac{\lambda}{2}\right)(|HH\rangle\langle VV| + |VV\rangle\langle HH|)\end{array}\right] \quad (S6)$$

The final state is then approximately given by

$$\rho_{out}(\lambda, \varepsilon) = \frac{K_0 \rho_{in}(\lambda,\varepsilon) K_0^\dagger}{\mathrm{Tr}\left[K_0 \rho_{in}(\lambda,\varepsilon) K_0^\dagger\right]} \quad (S7)$$

$$\approx \frac{1}{T_V + T_H \varepsilon^2}\left[\varepsilon^2 T_H |HH\rangle\langle HH| + T_V |VV\rangle\langle VV| + \varepsilon\sqrt{T_H T_V}\left(1 - \frac{\lambda}{2}\right)(|HH\rangle\langle VV| + |VV\rangle\langle HH|)\right]$$

where $K_0$ represents the Kraus operator of the metamaterial. We obtained the experimental parameters $\lambda_{exp}$ and $\varepsilon_{exp}$ summarized in Table 1 using the following relations given by the approximate form of the density matrices

$$\varepsilon_{exp} = \sqrt{\frac{\mathrm{Tr}[\rho_{exp}|HH\rangle\langle HH|]}{\mathrm{Tr}[\rho_{exp}|VV\rangle\langle VV|]}}, \quad (S8)$$

$$\lambda_{exp} = 2\left(1 - \frac{\mathrm{Tr}[\rho_{exp}|HH\rangle\langle VV|]}{\mathrm{Tr}[\rho_{exp}|HH\rangle\langle HH|]}\varepsilon_{exp}\right), \quad (S9)$$

where $\rho_{exp}$ represents the experimentally reconstructed density matrix in the distillation of partially mixed states. A similar derivation to the above can be performed for the non-maximally entangled partially mixed initial state for $|\Psi_\varepsilon\rangle$. Figure S3 shows all the density matrices of the initial partially mixed states and final distilled states that are summarized in

Table 1 in the main text. In the case where θ is small, the experimental density matrices have four dominant components as expected from the theoretical approximation. On the other hand, there are several additional components in the case of large θ.

**Figure S1. Characterization of metamaterials by quantum process tomography. Experimentally obtained process matrices (χ matrices) for the different metamaterials used in the experiment for entanglement distillation.** The process matrices are given in the basis defined by the single-qubit Pauli operators, $E_i =$ I, X, Y and Z, where a single qubit is modified as $\rho \to \sum_{ij} \chi_{ij} E_i \rho E_j^\dagger$. (a) Metamaterial sample 1 (left) and an ideal partial polarizer with $T_V = 0.11 \pm 0.01$ (right). The process fidelity of the sample to the ideal case is 0.93±0.01 (Tr(χ)=0.53±0.01). (b) Metamaterial sample 2 (left) and an ideal partial polarizer with $T_V = 0.13 \pm 0.01$ (right). The process fidelity of the sample to the ideal case is 0.92±0.02 (Tr(χ)=0.55±0.01). (c) Metamaterial sample 3 (left) and an ideal partial polarizer with $T_V = 0.16 \pm 0.01$ (right). The process fidelity of the sample to the ideal case is 0.95±0.01 (Tr(χ)=0.54±0.01). (d) Metamaterial sample 4 (left) and an ideal partial polarizer with $T_V = 0.21 \pm 0.01$ (right). The process fidelity of the sample to the ideal case is 0.94±0.01 (Tr(χ)=0.56±0.01). (e) Metamaterial sample 5 (left) and an ideal partial polarizer with $T_V = 0.27 \pm 0.01$ (right). The process fidelity of the sample to the ideal case is 0.92±0.02 (Tr(χ)=0.60±0.01). (f) Metamaterial sample 6 (left) and an ideal partial polarizer with $T_V = 0.41 \pm 0.01$ (right). The process fidelity of the sample to the ideal case is 0.90±0.01 (Tr(χ)=0.69±0.01). (g) Metamaterial sample 7 (left) and an ideal partial polarizer with $T_V = 0.69 \pm 0.02$ (right). The process fidelity of the sample to the ideal case is 0.87±0.02 (Tr(χ)=0.85±0.02).

**Figure S2. Distillation of highly entangled states from non-maximally entangled pure states using different metamaterial nanoantenna arrays.** (a) Density matrix of the initial state of the form $|\Phi_\varepsilon\rangle = (\varepsilon|H\rangle|H\rangle + |V\rangle|V\rangle)/\sqrt{1+\varepsilon^2}$. (b)-(h), Density matrices of the distilled states from each of the seven metamaterials in ascending order. Panel h corresponds to the density matrix shown in Figure 3 in the main text.

**Figure S3. Distillation of highly entangled states from non-maximally entangled mixed states using metamaterial nanoantenna arrays.** (a), (e), (i): Density matrices of the three starting mixed states of the form $\rho_{\varepsilon,\lambda} = \frac{1}{1+\varepsilon^2}\left[\varepsilon^2|HH\rangle\langle HH| + |VV\rangle\langle VV| + \varepsilon\left(1-\frac{\lambda}{2}\right)(|HH\rangle\langle VV| + |VV\rangle\langle HH|)\right]$. (b), (f), (j): Density matrices of the

distilled states for the starting mixed states of a,e,i . (c), (g), (k): Density matrices of the three starting mixed states of the form $\sigma_{\varepsilon,\lambda} = \frac{1}{1+\varepsilon^2}\left[\varepsilon^2|HV\rangle\langle HV| + |VH\rangle\langle VH| + \varepsilon\left(1-\frac{\lambda}{2}\right)(|HV\rangle\langle VH| + |VH\rangle\langle HV|)\right]$. (d), (h), (l): Density matrices of the distilled states for the starting mixed states of (c), (g), (k). See Table 1 in the main text for the estimated EOF, fidelity and purity of all the starting states and distilled states (left hand column for $\rho_{\varepsilon,\lambda}$ and right hand column for $\sigma_{\varepsilon,\lambda}$ in ascending order). The panels (i), (j), (k), (l) correspond to the density matrices shown in Figure 4 in the main text.

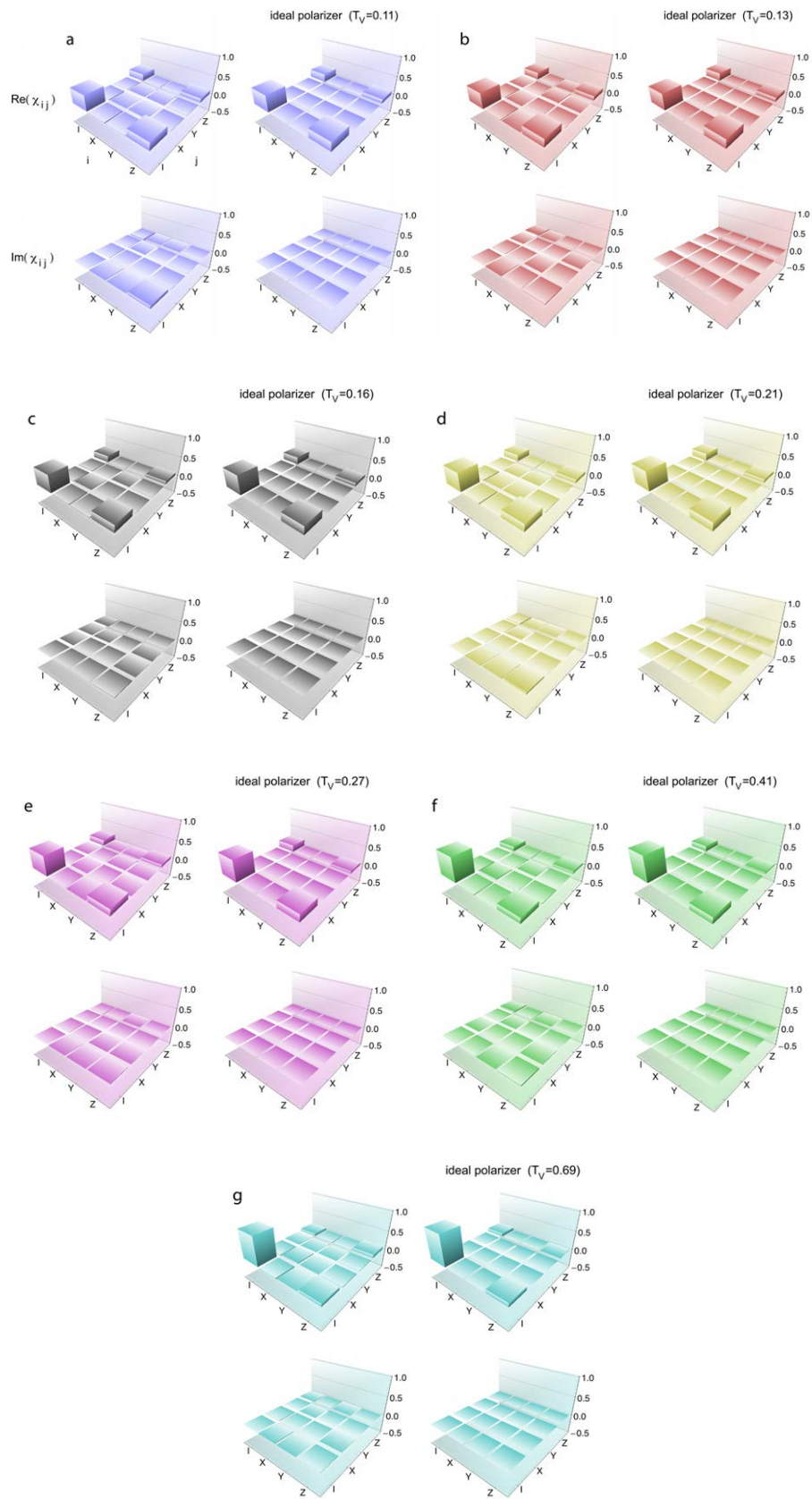

**Figure S1**

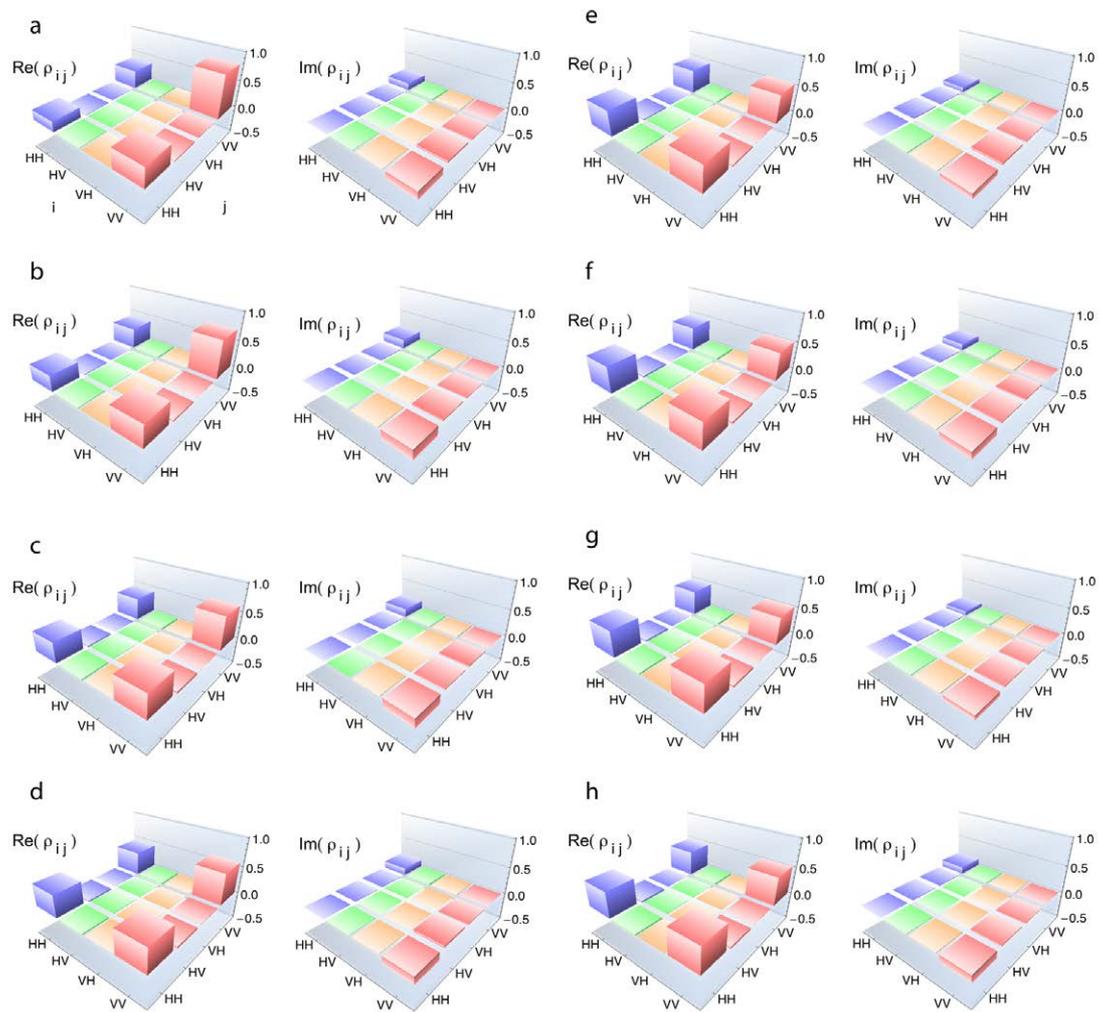

**Figure S2**

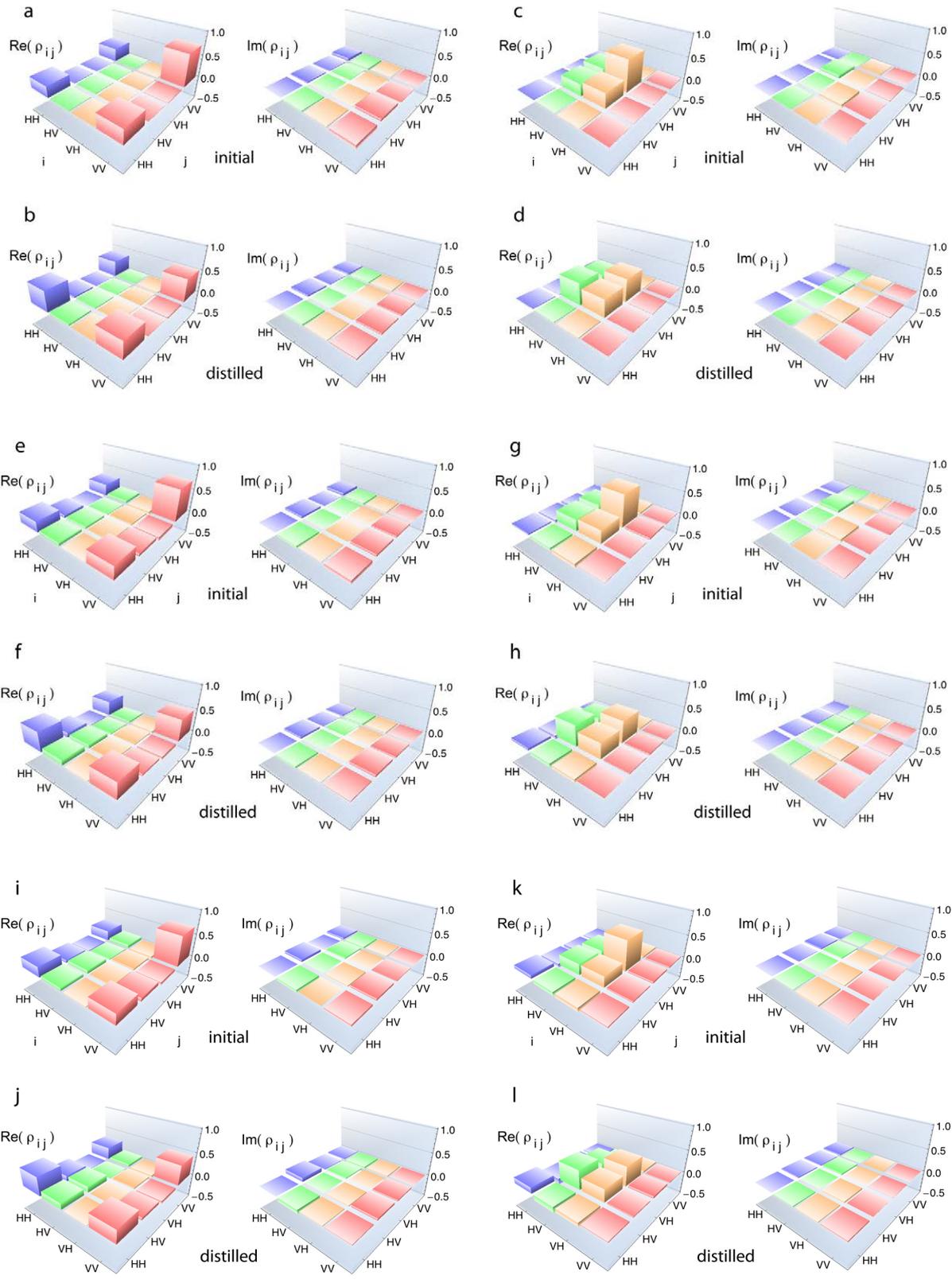

**Figure S3**